\documentclass[letterpaper]{elsart}

\usepackage{graphicx}
\journal{Mathematics and Computers in Simulation}

\newcommand{\D}{\mathrm d}
\newcommand{\E}{\mathrm e}
\newcommand{\I}{\mathrm i}

\begin{document}
\begin{frontmatter}
\title{Influence of the condensate and inverse cascade on the direct cascade in wave turbulence.}
\author[UNM,Landau]{A.\,O.~Korotkevich}
\ead{alexkor@math.unm.edu}
\address[UNM]{Department of Mathematics and Statistics, MSC03 2150, 1 University of New Mexico, Albuquerque, NM 87131-0001, USA}
\address[Landau]{L.\,D.~Landau Institute for Theoretical Physics RAS, 2
Kosygin Str., Moscow, 119334, Russian Federation}

\begin{abstract}
During direct numerical simulation of the isotropic turbulence of
surface gravity waves in the framework of Hamiltonian equations
formation of the long wave background or condensate was observed.
Exponents of the direct cascade spectra at the different levels of
an artificial condensate suppression show a tendency to become closer to the
prediction of the wave turbulence theory at lower levels of condensate.
A simple qualitative explanation of the mechanism of this phenomenon
is proposed.
\end{abstract}

\end{frontmatter}

\section{Introduction}
Description of the water waves appeals scientists attention during
several centuries. At the same time, first attempts to explain
observed spatial and temporal spectra are relatively recent.
One of the first and at the same time most famous works
is the paper by Phillips~\cite{Phillips1958} which
was, probably, the first attempt to give an explanation for power-like
spectra of surface gravity waves observed in numerous experiments.
In just one decade statistical theory of water waves, based on the
the kinetic equation for waves derived by Hasselmann~\cite{Hasselmann1962}
and solutions of this equation obtained from Zakharov's theory of wave (or weak)
turbulence~\cite{ZakharovPhD,ZFL1992}. These solutions~\cite{ZF1967,ZFL1992}
are stationary Kolmogorov solutions of the
kinetic equation corresponding to flux of energy from large to small
scales (direct cascade) and flux of wave action (waves ``number'')
from small to large scales (inverse cascade). Now the kinetic
equation is a base tool for wave forecasting.

The Hasselmann kinetic equation for waves was derived under some
assumptions, which include Gaussian
statistics for the wave field and prevalence of resonant
interactions~\cite{ZFL1992}. These assumptions are subject to
confirmation. Modern numerical methods allow us to perform wave field
modeling in the framework of kinetic equations faster than the real
processes in nature. At the same time, it is still not possible to create a
wave forecasting model based on direct numerical simulation of the primordial
dynamic equations. Fortunately, in practical applications we do not
need to know velocity and elevation at every point of the
surface. Statistics, especially mean wave hight and speed, are what
really matters for estimation of operational conditions of oil
platforms and cargo ships.  Such statistics are exactly the subject of the
theory of weak turbulence.  As a consequence, the problem of confirmation
and correction of the theory of wave turbulence is of great practical
importance.

Open sea observations confirm temporal and space spectra predicted
by the theory of wave turbulence~\cite{Toba1973,Donelan1985,Hwang2000}.
A comprehensive review
of experiments and comparison with the theory of the weak turbulence
can be found in~\cite{BPRZ2005,Zakharov2005,BBRZ2007}. At the same time
even the most recent, state of the art wave tanks experiments
give strange results~\cite{FFL2007,Lukaschuk2007}, demonstrating
dependence of the spectrum slope on the average steepness or
forcing level.

One of the ways from this deadend is direct numerical simulation
in the framework of dynamical equations.
Such numerical experiments can be separated in two major groups.
In swell simulation experiments initial condition is some spectral
distribution, there is no pumping. Temporal evolution of such initial condition
is simulated and than can be compared with corresponding simulation
in the framework of kinetic
equation~\cite{Tanaka2001,Onorato2002,Yokoyama2004,ZKPD2005,ZKPR2007,KPRZ2008}.
More interesting and more complex is the simulation of the
wave turbulence with pumping. Classical sandbox for such simulations
is an isotropic turbulence, which allows us to demonstrate of the features
of the wave turbulence and, at the same time, simplifies situation
due to additional smoothing of the spectra after angle averaging.
Although this situation looks pretty artificial,
the isotropic turbulence can be observed in the sea with large amount of
broken ice, for example, where waves create isotropic wavefield after
multiple reflections. But in such a case floating ice have to be taken into
account, which makes situation more complex.
Simulation of isotropic wave turbulence already confirmed
some prediction of the theory of weak turbulence. Direct cascade
spectra were obtained in~\cite{DKZ2003grav,DKZ2004,LNP2006}.
Inverse cascade spectra, although in simplified equations, were
obtained in~\cite{AS2006}. For the first time simultaneous
simulation of direct and inverse cascades for gravity waves in
the framework of primordial dynamical equations was performed
in~\cite{Korotkevich2008}.

Current paper is a continuation of the work~\cite{Korotkevich2008}.
The present article will concentrate on the condensate generation
and its influence on the direct cascade spectra of wave turbulence of
surface gravity waves. A more detailed and deep insight in the mechanism of the
condensate influence is presented. Starting from the results of
isotropic turbulence simulation author will explore in details
how presence of the condensate changes the slope of direct cascade spetrum.
It will be demostrated that the presence of the inverse cascade is also
important. At the same time in the pure situation with suppressed
condensate and inverse cascade observation shows spectrum slope close
to previous numerous results of simulations~\cite{DKZ2003grav,DKZ2004,LNP2006}.
A simple qualitative explanation of the mechanism of spectrum distortion is
provided. Direct analysis of the simulation results supports the theory.

\section{Theoretical background.}
Author follows previous works in theoretical description
of the system under
consideration~\cite{DKZ2003grav,DKZ2004,KPRZ2008,Korotkevich2008}.
\subsection{Dynamical equations.}
We study the potential flow of an ideal inviscid incompressible fluid with the velocity potential
$\phi=\phi(x,y,z;t)$
$$
\Delta \phi = 0,
$$
in the infinitely deep domain occupied by the fluid. Equations for the boundary conditions
at the surface are the following
\begin{equation}
\label{Laplas_boundary}
\begin{array}{c}
\displaystyle
\left.
\left(
\dot \eta + \phi'_x \eta'_x + \phi'_y \eta'_y\right)\right|_{z= \eta}
= \left. \phi'_z \right|_{z= \eta},\\
\displaystyle
\left. \left ( 
\dot \phi +
\frac{1}{2}\left|\nabla \phi \right|^2
\right ) \right |_{z= \eta} + g \eta = 0.
\end{array}
\end{equation}
Here $\eta(x,y;t)$ is the surface elevation with respect to still water, $g$ is the gravity
acceleration. Equations (\ref{Laplas_boundary}) are Hamiltonian~\cite{Zakharov1968}
with the canonical variables $\eta(x,y;t)$ and $\psi(x,y;t)=\phi(x,y,\eta(x,y;t);t)$
\begin{equation}
\label{Hamiltonian_equations}
\frac{\partial \eta}{\partial t} = \frac{\delta H}{\delta \psi}, \;\;\;\;
\frac{\partial \psi}{\partial t} = - \frac{\delta H}{\delta \eta},
\end{equation}
where $H$ is the Hamiltonian of the system
$$
H = \frac{1}{2} \int_{-\infty}^{+\infty} \D x \D y \left(
g\eta^2 + 
\int_{-\infty}^{\eta} |\nabla \phi|^2 \D z
\right),
$$
Unfortunately $H$ cannot be written in the close form as a functional of $\eta$ and $\psi$.
However one can limit Hamiltonian by first three terms of expansion in powers of $\eta$ and $\psi$
\begin{equation}
\label{Hamiltonian}
\begin{array}{l}
\displaystyle
H = H_0 + H_1 + H_2 + ...,\\
\displaystyle
H_0 = \frac{1}{2}\int\left( g \eta^2 + \psi \hat k  \psi \right) \D x \D y,\\
\displaystyle
H_1 =  \frac{1}{2}\int\eta\left[ |\nabla \psi|^2 - (\hat k \psi)^2 \right] \D x \D y,\\
\displaystyle
H_2 = \frac{1}{2}\int\eta (\hat k \psi) \left[ \hat k (\eta (\hat k \psi)) + \eta\Delta\psi \right] \D x \D y.
\end{array}
\end{equation}
Here $\hat k$ is a linear integral operator 
$\left(\hat k =\sqrt{-\Delta}\right)$, such that in $k$-space it corresponds to
multiplication of Fourier harmonics ($\psi_{\vec k} = \frac{1}{2\pi} \int \psi_{\vec r} e^{i {\vec k} {\vec r}} \D x \D y$)
by $k=\sqrt{k_{x}^2 + k_{y}^2}$. For gravity waves this
reduced Hamiltonian describes four-wave interaction.

Then dynamical equations (\ref{Hamiltonian_equations}) acquire the form
\begin{eqnarray}
\dot \eta = \hat k \psi - (\nabla (\eta \nabla \psi)) - \hat k [\eta
\hat k \psi] +\nonumber\\ + \hat k (\eta \hat k [\eta \hat k \psi]) +
\frac{1}{2} \Delta [\eta^2 \hat k \psi] + \frac{1}{2} \hat k [\eta^2
\Delta\psi] - \widehat F^{-1} [\gamma_k \eta_k],\nonumber\\ \dot \psi
= - g\eta - \frac{1}{2}\left[ (\nabla \psi)^2 - (\hat k \psi)^2
\right] - \label{eta_psi_system}\\ - [\hat k \psi] \hat k [\eta \hat k
\psi] - [\eta \hat k \psi]\Delta\psi - \widehat F^{-1} [\gamma_k
\psi_k] + \widehat F^{-1} [P_{\vec k}].\nonumber
\end{eqnarray}
Here dot means time-derivative, $\Delta$ --- Laplace operator, $\hat
k$ is a linear integral operator $\left(\hat k
=\sqrt{-\Delta}\right)$, $\widehat F^{-1}$ is an inverse Fourier
transform, $\gamma_k$ is a dissipation rate (according to recent
work~\cite{DDZ2008} it has to be included in both equations), which
corresponds to viscosity on small scales and, if needed, "artificial"
damping on large scales.  $P_{\vec k}$ is the driving term which
simulates pumping on large scales (for example, due to wind).  In the
$k$-space supports of $\gamma_{k}$ and $P_{\vec k}$ are separated by
the inertial interval, where the Kolmogorov-type solution can be
recognized.  These equations were derived as a result of Hamiltonian
expansion in terms of $\hat k \eta$.  From the physical point of view
the $\hat k$-operator is close to the derivative operator, so we
expand in powers of the slope of the surface. In most of the experimental
observations average slope of the open sea surface $\mu$ is of the
order of $0.1$, so such expansion is very reasonable.

These are dynamical equations which will be subject for modeling in a significant part of this article.
In order to understand the motivation of the paper we need to introduce statistical description
of the wave field.

\subsection{Kinetic equation.}
For the analisis of the wave field in is natural to use Fourier representation:
$$
\psi_{\vec k} = \frac{1}{2\pi} \int \psi_{\vec r} \E^{\I {\vec k} {\vec r}} \D^2 r,\;\;
\eta_{\vec k} = \frac{1}{2\pi} \int \eta_{\vec r} \E^{\I {\vec k} {\vec r}} \D^2 r.
$$
Wave field functions $\psi(\vec r, t)$ and $\eta(\vec r, t)$ are real valued functions,
so their Fourier transforms have Hermitian symmetry
$$
\psi_{\vec k} = \psi_{-\vec k}^*,\;\; \eta_{\vec k} = \eta_{-\vec k}^*.
$$
As a result we have two complex functions and each of them use effectively only half of
the wavenumbers plane. For further derivation is is convenient to introduce
so called normal variables:
\begin{equation}
\label{a_k_substitution}
a_{\vec k} = \sqrt \frac{\omega_k}{2k} \eta_{\vec k} + \I \sqrt \frac{k}{2\omega_k} \psi_{\vec k},\; \mbox{where   } \omega_k = \sqrt {g k}.\\
\end{equation}
In this variables for a wave field of one plain wave with wavevector $\vec k_0$ only harmonics $a_{\vec k_0}$
will be nonzero. We can state that we introduced elementary excitations of the system. Hamiltonian
equations in this variables has canonical form:
\begin{equation}
\label{Hamiltonian_eqs_canonical}
\dot a_{\vec k} = -\I \frac{\delta H}{\delta a_{\vec k}^{*}}\;\;\dot a^*_{\vec k} = \I \frac{\delta H}{\delta a_{\vec k}}.
\end{equation}
Hamiltonian in this terms allows us to separate different wave interaction processes:
\begin{equation}
\label{Ham_a_k}
\begin{array}{lll}
\displaystyle
H_0 = \int \omega_k |a_{\vec k}|^2 \D {\vec k},\\
\displaystyle
H_1 = \frac{1}{6}\frac{1}{2\pi}\int E_{\vec k_1 \vec k_2}^{\vec k_0} 
(a_{\vec k_1}a_{\vec k_2}a_{\vec k_0} + a_{\vec k_1}^{*}a_{\vec k_2}^{*}a_{\vec k_0}^{*})
\delta (\vec k_1 + \vec k_2 + \vec k_0) \D {\vec k_1}\D {\vec k_2}\D {\vec k_0} +\\
\displaystyle
+\frac{1}{2}\frac{1}{2\pi}\int C_{\vec k_1 \vec k_2}^{\vec k_0} 
(a_{\vec k_1}a_{\vec k_2}a_{\vec k_0}^{*} + a_{\vec k_1}^{*}a_{\vec k_2}^{*}a_{\vec k_0})
\delta (\vec k_1 + \vec k_2 - \vec k_0) \D {\vec k_1}\D {\vec k_2}\D {\vec k_0},\\
\displaystyle
H_2 = \frac{1}{4}\frac{1}{(2\pi)^2}\int W_{\vec  k_1 \vec k_2 \vec k_3 \vec k_4}
(a_{\vec k_1}a_{\vec k_2}a_{\vec k_3}a_{\vec k_4}
+ a_{\vec k_1}^{*}a_{\vec k_2}^{*}a_{\vec k_3}^{*}a_{\vec k_4}^{*})\times\\
\displaystyle
\times\delta (\vec k_1 + \vec k_2 + \vec k_3 + \vec k_4)\D {\vec k_1}\D {\vec k_2}\D {\vec k_3}\D {\vec k_4}+\\
\displaystyle
+\frac{1}{4}\frac{1}{(2\pi)^2}\int F_{\vec  k_1 \vec k_2 \vec k_3 \vec k_4}
(a_{\vec k_1}^{*}a_{\vec k_2}a_{\vec k_3}a_{\vec k_4}
+ a_{\vec k_1}a_{\vec k_2}^{*}a_{\vec k_3}^{*}a_{\vec k_4}^{*})\times\\
\displaystyle
\times\delta (\vec k_1 - \vec k_2 - \vec k_3 - \vec k_4)\D {\vec k_1}\D {\vec k_2}\D {\vec k_3}\D {\vec k_4}+\\
\displaystyle
+\frac{1}{4}\frac{1}{(2\pi)^2}\int D_{\vec  k_1 \vec k_2 \vec k_3 \vec k_4}
(a_{\vec k_1}a_{\vec k_2}a_{\vec k_3}^{*}a_{\vec k_4}^{*}
\delta (\vec k_1 + \vec k_2 - \vec k_3 - \vec k_4)\D {\vec k_1}\D {\vec k_2}\D {\vec k_3}\D {\vec k_4}.
\end{array}
\end{equation}
We omit formulae for matrix elements of these processes (they can be found in~\cite{KorotkevichPhD})
for the sake of conciseness.

For gravity waves on the surface of the deep fluid dispersion relation is $\omega=\sqrt{gk}$
and resonant conditions
\begin{eqnarray*}
\omega_{k_0} = \omega_{k_1} + \omega_{k_2},\;\;\; \vec k_0 = \vec k_1 + \vec k_2,
\end{eqnarray*}
are never fulfilled, which means that three-waves processes of decaying
and merging of the waves are prohibited. Such dispersion relations are of the ``non-decay type''.

In this case it is well known fact that cubic terms of the Hamiltonian (\ref{Ham_a_k})
can be eliminated by proper canonical transformation $a(\vec k,t) \longrightarrow b(\vec k,t)$.
Hamiltonian in new variables looks as follows
$$
\label{Ham_b_k}
\begin{array}{lll}
\displaystyle
H_0 = \int \omega_k |b_{\vec k}|^2 \D {\vec k},\\
\displaystyle
H_1 = 0,\\
\displaystyle
H_2 = \frac{1}{2}\frac{1}{(2\pi)^2}\int T_{\vec  k_1 \vec k_2 \vec k_3 \vec k_4}
b_{\vec k_1}^{*}b_{\vec k_2}^{*}b_{\vec k_3}b_{\vec k_4}
\delta (\vec k_1 + \vec k_2 - \vec k_3 - \vec k_4)\D {\vec k_1}\D {\vec k_2}\D {\vec k_3}\D {\vec k_4}.
\end{array}
$$
The formula of the matrix element $T_{\vec  k_1 \vec k_2 \vec k_3 \vec k_4}$ is too bulky
to give it here and can be found in~\cite{PRZ2003}.

If we suppose, that in our field phases and amplitudes are not correlated
one can introduce pair correlator
\begin{equation}
\langle a_{\vec k} a_{\vec k'}^*\rangle = n_{k} \delta (\vec k - \vec k').
\end{equation}
The $n_{\vec k}$ is measurable quantity, connected directly with observable correlation functions.
For instance, from $a_{\vec k}$ definition one can get
\begin{equation}
\label{I_k_expression}
I_k = \langle |\eta_{\vec k}|^2\rangle = \frac{1}{2}\frac{\omega_k}{g} (n_k + n_{-k}).
\end{equation}
For gravity waves we can also use the following correlation function
\begin{equation}
\langle b_{\vec k} b_{\vec k'}^*\rangle = N_{k} \delta (\vec k - \vec k').
\end{equation}
Because in the case of variables $b_{\vec k}$ all non-resonant processes are excluded
this is the correlation function for which kinetic equation is derived.

The relation connecting $n_k$ and $N_k$
is very simple (in the case of deep water)~\cite{Zakharov1999}
\begin{equation}
\frac{n_k - N_k}{n_k} \simeq \mu,
\end{equation}
where $\mu = (ka)^2$, here $a$ is a characteristic elevation of the free surface. In the case of the weak
turbulence $\mu \ll 1$.

The correlation function $N_k$ obeys the kinetic equation~\cite{Hasselmann1962, ZakharovPhD, ZFL1992}
\begin{equation}
\label{Kinetic_equation}
\frac{\partial N_k}{\partial t} = S_{nl}(N,N,N) + S_{in} (k) - S_{diss} (k),
\end{equation}
Here $S_{in}$ --- corresponds to input of energy, $S_{diss}$ --- dissipation of energy in the system, and
\begin{equation}
\begin{array}{l}
\displaystyle
S_{nl}(N,N,N)=4\pi \int \left| T_{\vec k,\vec k_1,\vec k_2,\vec k_3}\right|^2 \times\\
\displaystyle
\times(N_{k_1} N_{k_2} N_{k_3} +N_{k} N_{k_2} N_{k_3} - N_{k} N_{k_1} N_{k_2} -\\
\displaystyle
- N_{k} N_{k_1} N_{k_3})\delta (\vec k + \vec k_1- \vec k_2 - \vec k_3)
\delta (\omega_k + \omega_{k_1}- \omega_{k_2} - \omega_{k_3})
\D \vec k_1\D \vec k_2\D \vec k_3,\\
\end{array}
\end{equation}
describes nonlinear interaction of waves. Expression for $T_{\vec k,\vec k_1,\vec k_2,\vec k_3}$
can be found in~\cite{PRZ2003}.

\subsection{Kolmogorov-Zakharov solutions of kinetic equation.}
Let us consider stationary solutions of the kinetic equation assuming that
\begin{itemize}
\item The medium is invariant with respect to rotations;
\item Dispersion relation is a power-like function $\omega=a k^\alpha$;
\item $T_{\vec k,\vec k_1,\vec k_2,\vec k_3}$ is a homogeneous function:
$T_{\epsilon \vec k,\epsilon \vec k_1,\epsilon \vec k_2,\epsilon \vec k_3} = \epsilon^\beta T_{\vec k,\vec k_1,\vec k_2,\vec k_3}$.
\end{itemize}
Under these assumptions Zakharov~\cite{ZF1967,ZakharovPhD,ZZ1982,ZFL1992} obtained Kolmogorov-like solutions
corresponding to fluxes of two intergals of motion (energy and wave action or number of waves):
\begin{equation}
\begin{array}{l}
\displaystyle
n_k^{(1)} = C_1 P^{1/3} k^{-\frac{2\beta}{3} - d},\\
\displaystyle
n_k^{(2)} = C_2 Q^{1/3} k^{-\frac{2\beta - \alpha}{3} - d}.
\end{array}
\end{equation}
Here $d$ is a spatial dimension ($d=2$ in our case).
In the case of gravity waves on a deep water $\omega=\sqrt{gk}$ ($\alpha=1/2$) and, apparently, $\beta=3$.
As a result one can get:
\begin{equation}
\label{weak_turbulent_exponents}
\begin{array}{l}
\displaystyle
n_k^{(1)} = C_1 P^{1/3} k^{-4},\\
\displaystyle
n_k^{(2)} = C_2 Q^{1/3} k^{-23/6}.
\end{array}
\end{equation}
The first spectrum $n_k^{(1)}$ corresponds to the direct cascade, describing flux of energy from
large scales (pumping) to small scales (dissipation). The second spectrum $n_k^{(2)}$
describes to inverse cascade, corresponding to flux of number of waves (or wave action)
from small scales (pumping) to larger scales. In the case of swell or decaying turbulence
inverse cascade reveals itself as a downshift of spectral maximum to smaller wavenumbers, resulting
in growing long waves.

\subsection{Verification of the kinetic equation.}
As it was shown during derivation of kinetic equation, we neglect some processes,
such as multiple harmonics generation, because they are nonresonant. Also, statistical
description by definition does not take into account rare or catastrophic events, for
example solitons, freak wave formation, and wave breaking or whitecapping which
could be very important. It means that we need to verify these assumptions
using real life or numerical experiments.

Broad program of verification of kinetic equation for gravity waves was fulfilled
during the last decade. The key method of verification is numerical simulation in the
framework of primordial Hamiltonian equations for shape of the surface $\eta(\vec r, t)$
and velocity potential on the surface $\psi(\vec r, t)$. Such numerical experiments
provide complete information about the system, including phase and amplitude of
calculated function at every points of the surface at every moment of time.
Of course such luxury comes at a great cost of enormous computational complexity.
Due to rapid progress of computers available for computations and numerical algorithms
during last couple of decades such simulations became possible. The first work
which presented successful simulation of weak turbulence of gravity waves was
published in the beginning of 2000s~\cite{Onorato2002}. It was simulation of the decaying
turbulence. Decaying turbulence experiments start from initial spectral distribution
close to realistic swell which is propagating without pumping, decaying (loosing energy)
due to nonlinear transfer from spectral peak to the dissipative scale.
In the paper~\cite{Onorato2002} formation of weak turbulent Kolmogorov-Zakharov~\cite{ZFL1992}
spectral tail was confirmed. In the series of works~\cite{DKZ2003grav,KorotkevichPhD,DKZ2004}
the first confirmation of Kolmogorov-Zakharov spectra in systems with pumping
was provided as well as confirmation of independance of spectrum exponent from
the pumping force in some reasonable interval of input fluxes. Shortly after
that some possible obstacles for verification of Hasselmann equation through
wave tank experiments were pointed out~\cite{ZKPD2005}. In papers~\cite{KPRZ2008, ZKPR2007}
idea of modification of the dissipation term $S_{diss}$ in order to take additional
energy transfer to the dissipation scales due to whitecapping and wavebreaking was
developed through a series of massive numerical simulations of decaying turbulence.
In a recent paper~\cite{Korotkevich2008} some consequences of whitecapping
influence on a Kolmogorov-Zakharov spectra in wavetanks experiments were analyzed.

\section{Numerical simulation}
\subsection{Simulation parameters.}
Here we continue investigation of the problem which was started in~\cite{Korotkevich2008}.
We simulated the primordial dynamical
equations (\ref{eta_psi_system}) in a periodic spatial domain
$2\pi\times 2\pi$. In order to save computation time multigrid approach has been used.
Because formation of the inverse cascade is the most time consuming part of calculations
(characteristic nonlinear time is about $\omega_k/\mu \simeq 10\omega_k$ and $\omega_k \sim \sqrt{k}$),
we completed this calculation on the relatively small grid $256\times 256$. After that we gradually increased
resolution up to $1024\times 1024$ nodes. To check whether we loose something in the high frequency region
long time simulation on grid $256\times 256$ was performed. It demostrated absence of change in the pumping and high frequencies region.

\begin{figure}[htb]
\centering
\includegraphics[width=5.0in]{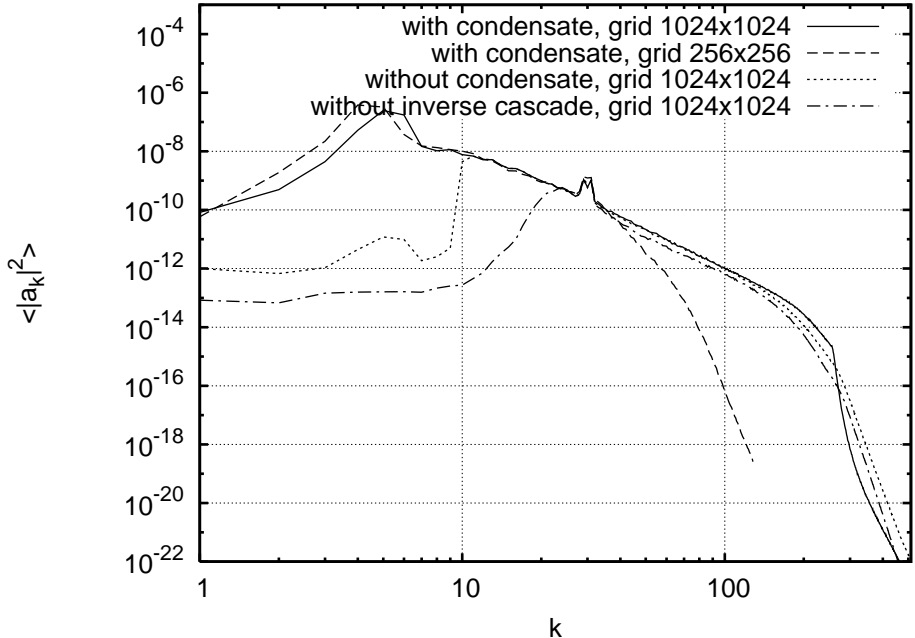}
\caption{\label{Spectra_all}Spectra $\langle|a_k|^2\rangle$. With
condensate on the $1024\times 1024$ grid (solid); on the $256\times
256$ grid with more developed condensate (dashed); without condensate
on the $1024\times 1024$ grid (dotted); without inverse cascade
on the $1024\times 1024$ grid (dashed-dotted).}
\end{figure}

  The used numerical code
was verified in~\cite{DKZ2003cap, DKZ2003grav, DKZ2004, ZKPD2005,
ZKPR2007, KPRZ2008}.  Gravity acceleration was $g=1$. The pseudo-viscous
damping coefficient had the following form
\begin{equation}
\gamma_{k} = \cases{
\gamma_0 (k - k_d)^2,\cr
0 - \mathrm{if}\;k \le k_d,\cr}
\end{equation}
where $k_d = 256$ and $\gamma_{0,1024} = 2.7\times10^{4}$ for the 
$1024\times 1024$ grid and $k_d = 64$ and $\gamma_{0,256} =
2.4\times10^{2}$ for the smaller $256\times 256$ grid. Pumping was an
isotropic driving force narrow in wavenumber space with random phase:
\begin{equation}
P_{\vec k} = f_k \E^{\I R_{\vec k} (t)}, f_k = \cases{
4 F_0 \frac{(k-k_{p1})(k_{p2}-k)}{(k_{p2} - k_{p1})^2},\cr
0 - \mathrm{if}\; k < k_{p1}\;\mathrm{or}\; k > k_{p2};\cr}
\end{equation}
here $k_{p1}=28,\; k_{p2}=32$ and $F_0 = 1.5\times 10^{-5}$; $R_{\vec
k}(t)$ was a uniformly distributed random number in the interval
$(0,2\pi]$ for each $\vec k$ and $t$. The initial condition was low
amplitude noise in all harmonics. Time steps were $\Delta t_{1024} =
6.7\times 10^{-4}$ and $\Delta t_{256} = 5.0\times 10^{-3}$.  We used
Fourier series in the following form:
\begin{eqnarray*}
\eta_{\vec k} = \widehat F[\eta_{\vec r}] = \frac{1}{(2\pi)^2}\int_{0}^{2\pi}\int_{0}^{2\pi}
\eta_{\vec r} \E^{\I \vec k\vec r}\D^2 r,\\
\eta_{\vec r} = \widehat F^{-1}[\eta_{\vec r}] = \sum\limits_{-N_x/2}^{N_x/2-1}\sum\limits_{-N_y/2}^{N_y/2-1}
\eta_{\vec k} \E^{-\I \vec k\vec r}.
\end{eqnarray*}
As it was reported in~\cite{Korotkevich2008} as a result
of simulation we observed formation of both direct and
inverse cascades (Fig.~\ref{Spectra_all}, solid line), although
exponents of power-like spectra were different from weak turbulent
solutions (\ref{weak_turbulent_exponents}). It is important to note
that development of the inverse cascade spectrum was arrested by the
discreteness of the wavenumber grid in agreement
with~\cite{DKZ2003cap, ZKPD2005, Nazarenko2006, NKO2008}. After that,
a large scale condensate started to form.
The reason of this phenomenon is the following.
It can be shown that resonant conditions for four waves interaction of gravity waves
are never fulfilled exactly on the discrete grid of wavevectors
(integer number for $2\pi\times 2\pi$ box):
$$
\omega_{k_1} + \omega_{k_2} - \omega_{k_3} - \omega_{k_4} \ne 0.
$$
Nonlinear frequency shift $\Gamma$ weakens resonant condition, now resonant curve has finite width:
$$
\omega_{k_1} + \omega_{k_2} - \omega_{k_3} - \omega_{k_4} \le \Gamma.
$$
Nonlinear frequency shift depends on $T_{\vec k_1 \vec k_2 \vec k_3 \vec k_4}$, which
is homogeneous function. It grows as $k^3$ when $k$ is increased. This is the reason why
it is much easier to get resonant interaction in the direct cascade region (high wavenumbers).
It also decreases as $k^3$ when $k \rightarrow 0$. It means that effectively grid become more
and more coarse at large scales.
It means that at some scale inverse cascade is stopped because four-waves
nonlinear interaction is ``turned off''.
At the same time flux of action still brings new waves to this scale.
We have ``condensation''~\cite{DF1996}
of waves. Similar processes were observed in 2D-hydrodynamics~\cite{Chertkov2007}
and plasma~\cite{Shats2007}

\begin{figure*}[htb]
\centering
\includegraphics[width=2.5in]{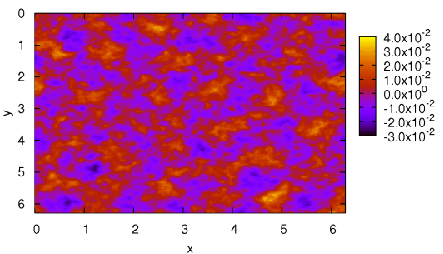}
\includegraphics[width=2.5in]{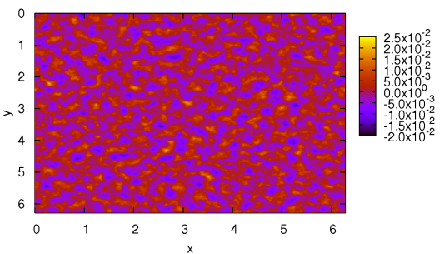}
\caption{\label{Surface}Surface of the fluid $\eta(\vec r)$ with
(left) and without (right) condensate.}
\end{figure*}

\begin{figure}[htb]
\centering
\includegraphics[width=2.5in]{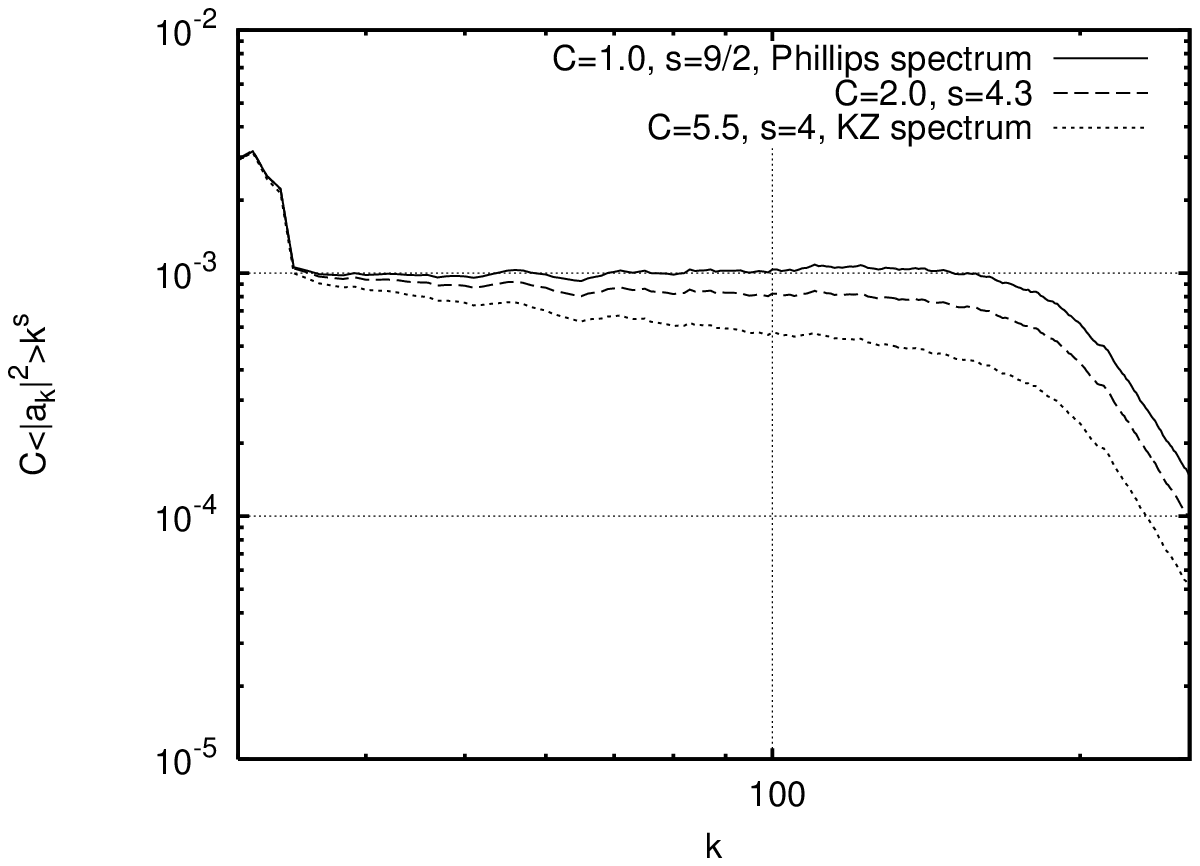}
\includegraphics[width=2.5in]{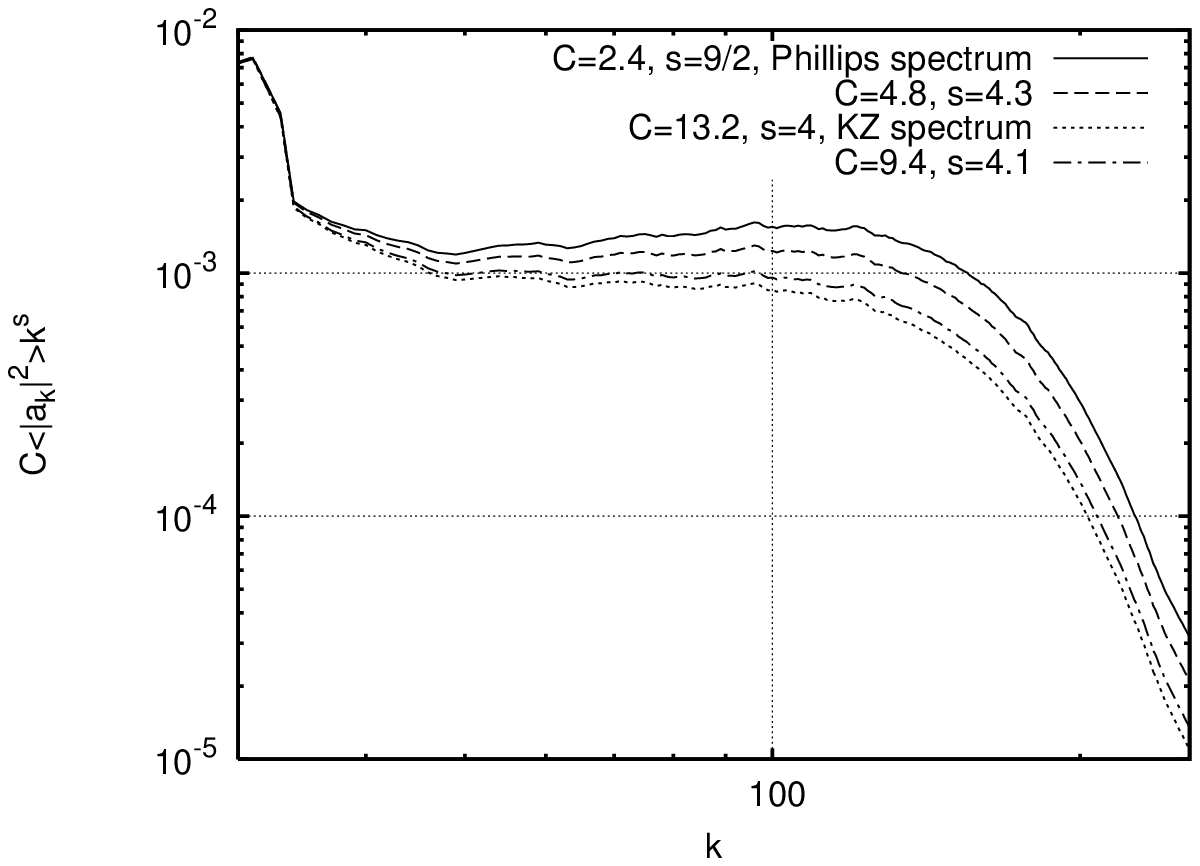}
\caption{\label{Compensated_spectra}Compensated direct cascade spectra
$C\langle |a_k|^2\rangle k^s$ with condensate and inverse cascade (left) and without them (right).}
\end{figure}

As one can see, the value of
wave action $|a_k|^2$ at the condensate region is more than an order
of magnitude higher than for the closest harmonic of the inverse
cascade. The dynamics of large scales became extremely slow after this
point. We managed to achieve downshift of the condensate peak by one step
of the wavenumber grid during long time simulation on a small grid 
($256\times 256$) (Fig.~\ref{Spectra_all}, line with long dashes). As
one can see, we observed elongation of the inverse cascade interval without
significant change of the slope. Unfortunately, the inertial interval for
inverse cascade was too short to exclude the possible influence of pumping
and condensate.
For direct cascade spectra we also used a log-log scale. Results are
present in Fig.~\ref{Compensated_spectra} (left). Formally, in this
case we have quite a long inertial interval $32 < k < 256$, but in
reality damping has an influence on the spectrum approximately up to
$k\simeq 180$. Still in this case we have more than half of a
decade. The theory of weak turbulence gives us dependence $\sim k^{-4}$
(\ref{weak_turbulent_exponents}), known as Kolmogorov-Zakharov
direct cascade spectrum for gravity waves. Nevertheless, one can see that we observe $k^{-9/2}$, known
as Phillips~\cite{Phillips1958,NZ2008} spectrum. Possible explanation
was provided in~\cite{Korotkevich2008}. The main idea was the following:
presence of the condensate stimulated additional whitecapping
and wavebraking processes by local increase of steepness. These processes
provide additional transfer of energy to the dissipation region. When this
dissipation is strong enough to balance flux of energy, such
mechanism should give Phillips spectrum as a result~\cite{NZ2008}.
It was shown, that suppression
of the condensate (see Figure~\ref{Surface})
changes spectrum exponent to some value closer to
the theory of weak turbulence. However, at that time we failed to avoid
long waves' influence on the direct cascade spectrum completely.
The longest waves of inverse cascade were providing long wave background
which influence distorted the spectrum.

In order to eliminate this influence let us suppress inverse cascade completely
by artificial damping in low wavenumbers. The result of simulation is shown
in Figure~\ref{Compensated_spectra} (right panel).

For comparison we have put in the same figure initial spectrum for direct cascade (left).
As one can see, the exponent of the
spectrum changed and now almost coincides with the prediction of the weak turbulent
theory. The tiny difference may be a result of the influence of the
left peak, which is the result of insufficient length of damping region. At the same time,
longer artificial damping can start to influence the pumping area and change the input flux
significantly.

\begin{figure}
\centering
\includegraphics[width=5.0in]{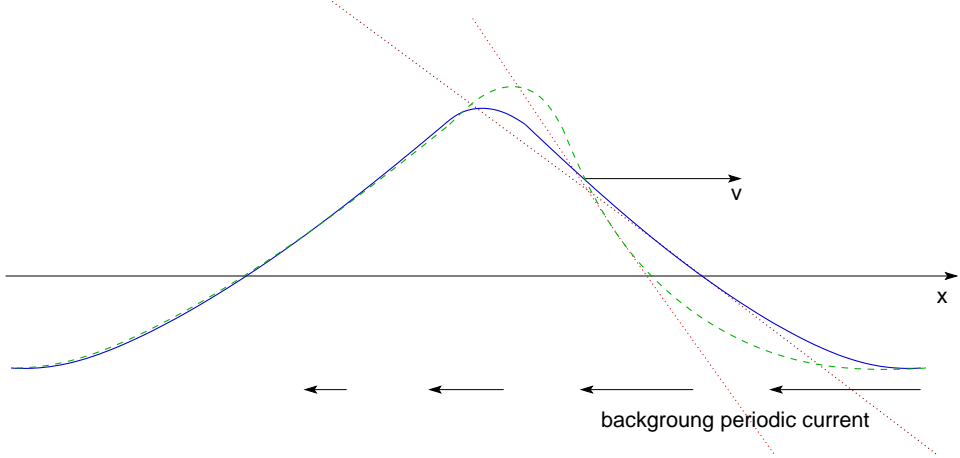}
\caption{\label{Steepening} Initial wave (solid line) become steeper (dashed line) in the result
of interaction with the backround current.}
\end{figure}

\begin{figure}[hbt]
\centering
\includegraphics[width=5in]{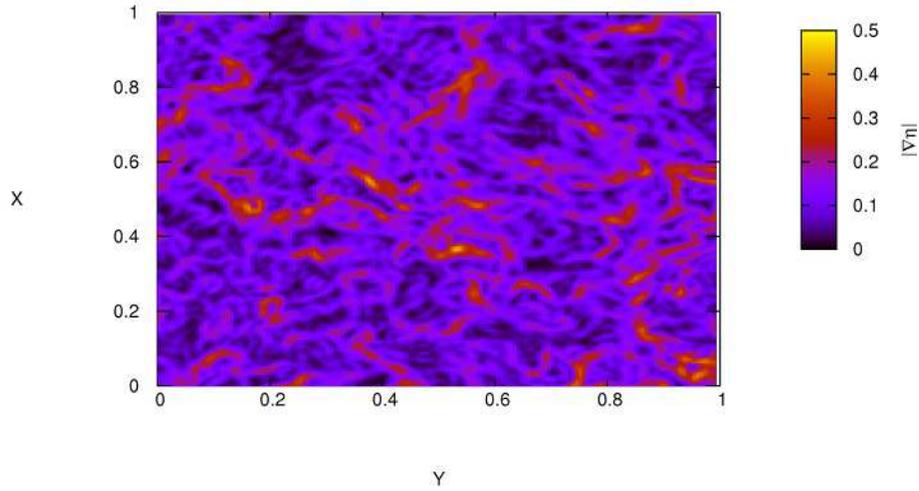}
\caption{\label{Grad_Condensate} Gradient of some part of the surface. In the presence of condensate.}
\end{figure}

\begin{figure}[hbt]
\centering
\includegraphics[width=5in]{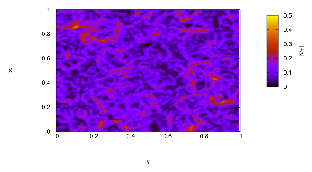}
\caption{\label{Grad_NoCondensate} Gradient of some part of the surface. In the absence of condensate.}
\end{figure}

\begin{figure}[hbt]
\centering
\includegraphics[width=5in]{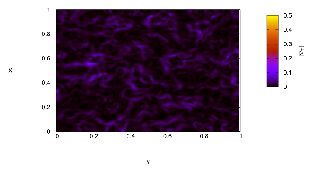}
\caption{\label{Grad_NoInv} Gradient of some part of the surface. In the absence of inverse cascade.}
\end{figure}

A qualitative explanation of the condensate's influence on the short
waves could be the following: let us consider a propagating 
wave with some given slope at its front; a much longer wave can be
treated as a presence of a strong background flow. If the direction of
the flow is opposite to direction of the wave's propagation, the slope of the
wave's front will increase (see Figure~\ref{Steepening}).
This is what we see in our
simulations. The average steepness $\mu = \sqrt{\langle|\vec
\nabla\eta|^2\rangle}$ has changed: with condensate $\mu_{c}\simeq
0.14$ (Figure~\ref{Grad_Condensate}),
without condensate $\mu_{nc}\simeq 0.12$ (Figure~\ref{Grad_NoCondensate}),
after suppression of inverse cascade $\mu_{nc}\simeq 0.1$ (Figure~\ref{Grad_NoInv}).
We should mention that very high values of gradient of the surface (higher that
value for a limiting Stokes wave) are possible in our model only due to very
strong damping at high wavenumbers and relatively narrow inertial interval, which
prevents waves from real breaking by strong dissipation of energy for every breaking
onset. We regularize our model by this dissipation. At the same time here we don't need
to know details of the energy dissipation (how in details wave breaks and produce
very short waves), all we need is to dissipate all the energy involved in the breaking.

\section{Conclusion}
The observed phenomena of condensate formation and its influence on
the direct cascade have to be quite common for experiments in
experimental wave tanks. Finite size of the surface area
gives us the same discrete grid of wavevectors as in the case
of periodic boundary conditons. It means that at some large scale
four waves nonlinear interaction will be arrested, which provides
conditions for condensate formation. Such condensate will distort
the direct cascade spectrum, changing its exponent through stimulation
of whitecapping, resulting in additional dissipation. It is worth
to note that this is threshold phenomenon with respect to average steepness
as it was shown in~\cite{BBY2000}. If such
dissipation is strong enough to balance the flux of energy,
following~\cite{NZ2008} we shall get Phillips spectrum, as it was
demonstrated in the present paper. Gradually suppressing condensate and inverse
cascade one can achieve spectrum corresponding to the stationary
solution of the kinetic equation. Simple mechanism of condensate
interaction with short waves of direct cascade was proposed.
This interaction  along with other distortion possibilities
(see, for example, \cite{Korotkevich2008D}) can be the reason
of observed spectra deviation from the weak turbulence predictions.
These results can be used for explanation of experimental data
obtained in the wavetanks.

\section{Acknowledgments}
The author would like to thank V.\,E.~Zakharov, V.\,V.~Lebedev, and I.\,V.~Kolokolov for fruitful discussions.

This work was partially supported by RFBR grant 06-01-00665-a,
the Program ``Fundamental problems of nonlinear dynamics'' from the RAS
Presidium and ``Leading Scientific Schools of Russia'' grant
NSh-7550.2006.2.

The author would also like to thank the creators of the
free open-source fast Fourier transform library FFTW~\cite{FFTW}
for this fast, cross platform, and reliable software.

\end{document}